\documentclass[10pt]{iopart}
\usepackage{graphicx}
\usepackage{xcolor}
\pdfoutput=1
\usepackage{iopams} 

\eqnobysec
\newcommand{\beq}{\begin{equation}}
\newcommand{\beqa}{\begin{eqnarray}}
\newcommand{\eeq}{\end{equation}}
\newcommand{\eeqa}{\end{eqnarray}}

\renewcommand{\d}{{\rm d }}

\renewcommand{\c}{{\cal C}}

\newcommand{\s}{{\sigma}}

\newcommand{\w}{{\bar w}}

\newcommand{\eq}{\mathrm{eq}}

\newcommand{\stat}{\mathrm{stat}}
\newcommand{\asym}{\mathrm{asym}}
\newcommand{\rel}{\mathrm{relax}}

\newcommand{\al}{a}

\begin{document}

\title{Dynamics of the two-dimensional directed Ising model in the paramagnetic phase}
\date{\today}
\author{C Godr\`eche$^1$ and M Pleimling$^2$}
\address{
$^1$Institut de Physique Th\'eorique, CEA Saclay and CNRS\\
91191 Gif-sur-Yvette cedex, France}\smallskip
\address{$^2$Department of Physics,
Virginia Polytechnic Institute and State University
Blacksburg, Virginia 24061-0435, USA
}
\begin{abstract}
We consider the non-conserved dynamics of the Ising model on the two-dimensional square lattice, where each spin is influenced preferentially by its East and North neighbours.
The single-spin flip rates are such that the stationary state is Gibbsian with respect to the usual ferromagnetic Ising Hamiltonian.
We show the existence, in the paramagnetic phase, of a dynamical transition between two regimes of violation of the fluctuation-dissipation theorem in the nonequilibrium stationary state: a regime of weak violation where the stationary fluctuation-dissipation ratio is finite, when the asymmetry parameter is less than a threshold value, and a regime of strong violation where this ratio vanishes asymptotically above the threshold.
The present study suggests that this novel kind of dynamical transition in nonequilibrium stationary states, already found for the directed Ising chain and the spherical model with asymmetric dynamics, might be quite general. 
In contrast with the later models, the equal-time correlation function for the two-dimensional directed Ising model depends on the asymmetry.
\end{abstract}

\maketitle

\section{Introduction}

Physical systems which are defined by dynamical rules are not in general time reversible.
Their stationary states can be arbitrarily difficult to determine and there is no general prescription to this end.
Conversely, if the dynamics fulfills the condition of detailed balance with respect to the Hamiltonian describing the statics of the model, or, in other words, is reversible, then the system reaches an equilibrium state, described by statistical physics.
Between these two cases lies an intermediate situation, where only the weaker constraint of global balance is imposed on the dynamics, in the sense that the stationary master equation of the process is globally satisfied by the same measure as at equilibrium.
One may then wonder whether there exist models depending continuously on a parameter (or more generally on a set of parameters) such that the dynamics is either reversible or irreversible, in the later case with global balance being satisfied, according to the value of this parameter.
If so, it is to be expected that the dynamical properties of these models will depend on whether the dynamics is reversible or not.

The directed Ising chain with non conserved dynamics~\cite{gb2009,cg2011,cg2013} is a prototypical example of this kind of models.
Directedness means that the flipping spin is unequally influenced by its two neighbours.
For the linear chain, asymmetry of the dynamics and irreversibility go hand in hand:
for any choice of the flipping rate function, invariant under spin reversal symmetry, the system relaxes to a stationary state described by the Boltzmann-Gibbs measure corresponding to the equilibrium Hamiltonian of the system, with the proviso that the parameters entering the rate function obey a condition fixing the temperature~\cite{cg2013}.
For special choices of these parameters the dynamics is reversible; restricting furthermore this choice leads to the solvable Glauber dynamics~\cite{glau}.
More generally, a one-parameter family of asymmetric dynamics
can been shown to keep the exact solvability of the symmetric Glauber dynamics,
allowing a detailed analysis of its dynamical features~\cite{cg2011}.

Such a Gibbsian asymmetric kinetic model,
where the flipping spin is not equally influenced by its two neighbours, therefore
provides a benchmark to study the phys\-ical consequences of irreversibility, in particular the properties of the resulting non\-equi\-li\-brium stationary state.
One of the most striking results is the existence of two regimes
of violation of the fluctuation-dissipation theorem in the non\-equi\-li\-brium stationary state.
Indeed, the fluctuation-dissipation theorem~\cite{fdt}, which gives a relation between response and correlation, is violated at stationarity as soon as the asymmetry parameter is non zero, because the dynamics becomes irreversible.
Surprisingly this violation depends on the strength of the asymmetry: a dynamical transition is observed between a regime of weak violation with a finite asymptotic stationary fluctuation-dissipation ratio (the limit at large times of the ratio between the stationary response and correlation functions) when the asymmetry parameter is less than a threshold value, and a regime of strong violation
where this ratio vanishes asymptotically above the threshold~\cite{cg2011}.
In the low temperature scaling regime, the asymptotic transient fluctuation-dissipation ratio depends on the strength of the asymmetry~\cite{cg2011}, and in particular vanishes at long times, as soon as the asymmetry is present, while in the absence of asymmetry it takes a universal value equal to $1/2$~\cite{gl2000,lipp2000}.
Another example where the same phenomena occur is the spherical model under asymmetric Langevin dynamics~\cite{gl2013}.

In the present work we address the situation of the directed Ising model on the two-dimensional square lattice.
We consider the case where the flipping spin is more influenced by its East and North neighbours than by its West and South ones.
As shown in~\cite{gb2009,cg2013} there exist rates such that, though the dynamics is asymmetric and irreversible, the stationary measure is still given by the Boltzmann-Gibbs distribution.
Our aim is to investigate how the dynamical properties of the model are modified in the presence of the asymmetry and in particular whether the properties found in one dimension have a natural generalization in two dimensions.
The present study is devoted to the paramagnetic phase:
{\it our purpose is to investigate how the system, prepared in a disordered initial condition, relaxes to its stationary state, then fluctuates in this stationary state}.
In a companion paper~\cite{ustocome}, we shall address the situation encountered at criticality and in the ferromagnetic phase.

We find that the two regimes of violation of the fluctuation-dissipation theorem uncovered for the linear chain are still present.
This can be traced back, as was already the case of the chain, to the different behaviours of the linear and non linear relaxation rates: these two rates, initially equal in the absence of an asymmetry, take different values above a threshold value of the asymmetry parameter.
We also find that the equal-time correlation function bears a dependence on the strength of the asymmetry, which reflects the fact that the dynamics of the two-dimensional Ising model is non linear, in contrast with the situation prevailing for the one-dimensional chain ot the spherical model, where the correlation function is independent of the asymmetry parameter, in line with the fact that the dynamical equations are linear.
However, this dependence of the equal-time correlation function on the strength of the asymmetry is very weak in the paramagnetic phase.
At low temperature it becomes much stronger~\cite{ustocome}.

Let us finally point out an important difference between directed Ising models and other nonequilibrium models, such as the 
zero-range process~\cite{spitzer}, or the Ising chain evolving under Kawasaki dynamics and submitted to an external drive (the so-called KLS model~\cite{kls}).
These two models share, to a certain extent, similarities with the directed Ising models.
Indeed, the stationary measure of the zero-range process, where particles hop from site to site on a graph with a rate only depending on the occupation of the departure site, is independent on whether the dynamics is reversible or not~\cite{zrp}.
The same holds for the KLS model, in the sense that, with an appropriate choice of rates, the system relaxes towards a stationary state, the measure of which is identical to the equilibrium Boltzmann-Gibbs measure of the undriven chain~\cite{kls,lg2006,cg2013}.
However, in both examples, the dynamics is irreversible because there is an applied field on the system, which implies that there is transport of mass along the system, while for directed Ising models the asymmetry does not result from an externally applied field, hence {\it there is no macroscopic current transported in the system}.

The outline of the paper is as follows.
Section~\ref{sec:def} is devoted to the definition of the dynamical rules of our model.
In section~\ref{sec:remind} we give a brief reminder of the main results of the study performed in~\cite{cg2011} on the case of the linear chain, which will serve as a template for the study of the two-dimensional model.
Sections~\ref{sec:bulk1} and~\ref{sec:bulk2} are the main parts of the present work.
We first investigate the role of the asymmetry on the temporal evolution of the equal-time correlation function.
We then demonstrate, by means of extensive numerical simulations, the existence of two regimes of violation of the fluctuation-dissipation theorem.
We conclude by a discussion of our results.

\section{Definition of the dynamical rules of the directed Ising model}
\label{sec:def}

In this section we give the definition of the dynamical rules of our model.

\subsection{Expression of the rates with asymmetric dynamics}

We consider a system of $N$ Ising spins on a two-dimensional square lattice, with sites labeled $i=1,\ldots, N$.
The energy of the configuration $\c=\{\s_1,\ldots,\s_N\}$ reads
\beqa\label{hamilt}
E(\c)=- J\sum_{(i,j)}\s_i\s_j,
\eeqa
where $(i,j)$ are nearest neighbours and $J>0$ is the coupling constant.
From now on we will set $J=1$ and $k_B=1$ and denote the reduced coupling constant by $K=1/T$.

At each instant of time, a spin, denoted by $\s$, is chosen at random and flipped with a rate $w(\s;\{\s_E,\s_N,\s_W,\s_S\})$, denoted for short by $w(\s;\{\s_j\})$, where $\{\s_j\}=\{\s_E,\s_N,\s_W,\s_S\}$ are the neighbours of the central spin $\s$, and E, N, W, S stand for East, North, etc.\footnote{The notation $\s_N$, where $N$ stands for North, should not be confused with the notation for the spin with index $N$, size of the system.}
We choose the following form of the rate function:
\beqa
w(\s;\{\s_j\})&=&
\frac{\alpha}{2}\big[
1-\gamma\frac{1+V}{2}\s(\s_E+\s_N)-\gamma\frac{1-V}{2}\s(\s_W+\s_S)
\nonumber\\
&+&\gamma^2\frac{1+V}{2}\s_E\s_N+\gamma^2\frac{1-V}{2}\s_W\s_S
\big],
\label{rate}
\eeqa
where $\alpha$ fixes the scale of time, $\gamma=\tanh 2K$ and $V$ is the asymmetry parameter, or velocity, which allows to interpolate between the symmetric case ($V=0$) and the totally asymmetric ones ($V=\pm1$).
As commented in the next section, this expression of the rate function satisfies the condition of global balance, or otherwise stated, leads to a Gibbsian stationary measure with respect to the Hamiltonian~(\ref{hamilt}).
It represents one, among many, possible expression of a rate function, for the kinetic Ising model on the square lattice, possessing this property.
This expression is invariant under up-down spin symmetry.
In~(\ref{rate}), for $V>0$ (resp. $V<0$) the couple (N, E) (resp. (S, W)) is more influential on the central spin than the other one. 

The two particular cases of symmetric dynamics ($V=0$), and completely asymmetric dynamics ($V=\pm1$), deserve special attention.
First, if $V=0$, then the rate function
\beqa
w(\s;\{\s_j\})=\frac{\alpha}{2}\big[
1-\frac{\gamma}{2}\s(\s_E+\s_N+\s_W+\s_S)
+\frac{\gamma^2}{2}(\s_E\s_N+\s_W\s_S)
\big]
\nonumber\\
\label{rate0}
\eeqa
satisfies the condition of detailed balance.
This form is different from the usual Glauber rate function (see~(\ref{glau2D}) and~(\ref{eq:glauber}) below).
%Yet it yields reversible dynamics.
It does not possess all the symmetry of the former inasmuch as all neighbouring spins of the central spin do not play equivalent roles, i.e., cannot be interchanged (see also the comment at the end of this section).
This can be demonstrated by looking at the values taken by $w(\s;\{\s_j\})$ according to the values of the local field $\s_E+\s_N+\s_W+\s_S$: if this sum is equal to $\pm2$ then all rates are the same.
In contrast, if the sum vanishes then the rate function takes two different values, $\alpha(1\pm\gamma^2)$, according to the configuration of the neighbours $\{\s_E,\s_N,\s_W,\s_S\}$: $\alpha(1+\gamma^2)$ corresponds to $\{++--\}$ or $\{--++\}$, while $\alpha(1-\gamma^2)$ corresponds to all other configurations.
This is a manifestation of an anisotropy in the dynamics.
For the Glauber case, 
\beq\label{glau2D}
w(\s;\{\s_{j}\})=\frac{\alpha}{2}\big[1-\s\tanh K(\s_E+\s_N+\s_W+\s_S)\big],
\eeq
if the sum vanishes, the rate function takes only one value, $\alpha/2$.
Therefore, at zero temperature ($\gamma=1$), all configurations such that the local field vanishes have a non-zero rate in the Glauber case, while in our model only the two configurations of neighbours $\{++--\}$ or $\{--++\}$ lead to a potential flip of the central spin.

The case with $V=1$ corresponds to the totally asymmetric dynamics where the central spin $\s$ is influenced by its East and North neighbours only.
Then
\beq\label{rate:ne}
w(\s;\{\s_j\})=\frac{\alpha}{2}
\left[1-\gamma\s(\s_E+\s_N)+\gamma^2\,\s_E\s_N\right].
\eeq
A similar expression, involving $\s_W$ and $\s_S$, holds for $V=-1$:
\beq\label{rate:sw}
w(\s;\{\s_j\})=\frac{\alpha}{2}
\left[1-\gamma\s(\s_W+\s_S)+\gamma^2\,\s_W\s_S\right].
\eeq
A remarkable fact about these expressions is that they are {\it unique}, up to the scale of time fixed by the choice of $\alpha$: the rate function is uniquely determined by the condition of global balance, when only two neighbours, chosen among the four possible ones, are influential upon the central spin~\cite{gb2009,cg2013}.

As a final remark, let us point out that the rate function~(\ref{rate}) is a linear combination of~(\ref{rate:ne}) and of~(\ref{rate:sw}), which makes its definition rather natural.
Of course, one could define a more symmetrical rate function by taking a linear superposition of the expressions corresponding to the four couples NE~(\ref{rate:ne}), NW, SW~(\ref{rate:sw}), and SE.
This choice would however imply the presence of more than one parameter in the rate function.

\subsection{Derivation of the rate function~(\ref{rate})} 

With the aim of understanding the origin of the expression~(\ref{rate}), we briefly come back on the main steps of the method used in~\cite{gb2009,cg2013} to derive a set of constraints between the transition rates that need
to be satisfied in order for the stationary state measure to be Gibbsian. 

The dynamics consists in flipping the spin $\s$, chosen at random, with a rate 
$w(\overline{\c}|\c)$, which is a more formal notation for $w(\s;\{\s_j\})$, corresponding to the transition between configurations $\c$ and 
$\overline{\c}=\{\s_1,\ldots,-\s_{},\ldots,\s_{N}\}$.
We choose periodic boundary conditions.
At stationarity, the master equation expresses that losses are equal to gains, and reads
\beq\label{master0}
P(\c)\sum_{\overline{\c}}w(\overline{\c}|\c)=\sum_{\overline{\c}}w(\c|\overline{\c})P(\overline{\c}),
\eeq
where, by hypothesis, the weight of the configuration $\c$ reads
\beq
P(\c)\propto\e^{-E(\c)/T}.
\eeq
After division of both sides by the weight $P(\cal C)$,
eq.~(\ref{master0}) can be rewritten as
\beq\label{mast}
\sum_{\overline{\c}}\big( w(\overline{\c}|\c)-w(\c|\overline{\c})\e^{-\Delta E/T}\big)=0,
\eeq
where the change in energy due to the flip reads
\beq\label{del}
\Delta E=E(\overline{\c})-E(\c)=2\sigma (\sigma_E+\sigma_N+\sigma_W+\sigma_S).
\eeq

While detailed balance consists in equating each individual term appearing on both sides of the stationary master equation~(\ref{mast}), global balance means that this equation is satisfied as a whole.
In other words~(\ref{mast}) expresses the global balance condition on the rates, which are the unknown quantities of the latter.
It leads to a set of constraints on the rates which should be satisfied in order for the stationary state to be Gibbsian with Hamiltonian~(\ref{hamilt}).
(See~\cite{gb2009,cg2013} for details.)

There are $32$ ($2^5$) possible configurations of the group of spins $\{\s;\s_{E},\s_{N},\s_W,\s_S\}$ involved in the flipping of the central spin $\s$.
Therefore, in the absence of any constraints on the dynamics, the most general form of the rate function~$w(\s;\{\s_j\})$ takes 32 values,
and thus depends on 32 parameters.
We hereafter consider the simpler case where the rate function is invariant under spin reversal, which is possible in the absence of a magnetic field, i.e., we request the rate function to be invariant under the change of sign of $\s$, $\s_E,\s_N, \s_W, \s_S$:
\beq\label{symm}
w(-\s;\{-\s_E,-\s_N,-\s_W,-\s_S\})=w(\s;\{\s_E,\s_N,\s_W,\s_S\}).
\eeq

The number of possible values of the rate function is therefore halved and is equal to the number of different environments of the central spin $\s$, i.e., of configurations of its neighbours.
There are $16$ ($2^4$) such configurations, labelled by the index $\al$.
We denote the 16 rates with $\s=+1$ by $w_{\al}$ and the other $16$ rates, corresponding to $\s=-1$, by $\bar w_{\al}$:
\beq\label{wal}
w_{\al}=w(\s=+1;\{\s_j\}_\al),\qquad \bar w_\al=w(\s=-1;\{\s_j\}_\al).
\eeq
The latter are obtained from the former by the spin symmetry relation~(\ref{symm}), 
yielding
\beq\label{wsymm}
\bar w_\al=w_{17-\al}, 
\eeq
(see Table~\ref{tab:2d}).
As a result, the number of independent parameters is decreased to 16.

Imposing now the global balance condition~(\ref{mast})
yields, as shown in~\cite{gb2009,cg2013}, the set of constraints
\beqa\label{six}
\e^{8 K}\, w_{1}-\w_{1}=0,\nonumber \\
w_{6}-\w_{6}=0,\nonumber \\
\e^{4 K}\, w_{2}-\w_{2}+\e^{4 K}\, w_{5}-\w_{5}=0,\nonumber \\
\e^{4 K}\, w_{3}-\w_{3}-(w_{8}-\e^{4 K}\, \w_{8})=0,\nonumber \\
\e^{4 K}\,w_2-\w_2-(\e^{4 K}\,w_3-\w_3)
+\frac{2\e^{4 K}}{1+\e^{4 K}}(w_{7}-\w_{7})=0,\nonumber\\
\e^{4 K}\,w_2-\w_2+\e^{4 K}\,w_3-\w_3-\frac{2\e^{4 K}}{1+\e^{4 K}}(w_{4}-\w_{4})=0.
\eeqa

\begin{table}[ht]
\caption{List of local configurations and corresponding values of the rate function for the 2D square lattice.
There are 16 possible rates $w_{\al}$, with $\s=+1$, corresponding to the 16 possible configurations $\{\s_j\}$, labelled by $\al$, of the four neighbours of the central spin, taken in the order: East, North, West, South.
The 16 remaining rates $\w_\al$, with $\s=-1$, are deduced from the former, due to the spin symmetry (see~(\ref{wsymm})).}
\label{tab:2d}
\begin{center}
\begin{tabular}{|c||c|c||c|c|}
\hline
$\al$&$\s;\{\s_j\}$&$w_\al$&$\s;\{\s_j\}$&$\w_\al$\\
\hline
$1$&$+;{++++}$&$w_{1}$&$-;++++$&$\w_1=w_{16}$\\
$2$&$+;{+++-}$&$w_2$&$-;+++-$&$\w_2=w_{15}$\\
$3$&$+;{++-+}$&$w_3$&$-;{++-+}$&$\w_3=w_{14}$\\
$4$&$+;{++--}$&$w_4$&$-;++--$&$\w_4=w_{13}$\\
$5$&$+;+-++$&$w_{5}$&$-;+-++$&$\w_5=w_{12}$\\
$6$&$+;+-+-$&$w_{6}$&$-;+-+-$&$\w_6=w_{11}$\\
$7$&$+;+--+$&$w_{7}$&$-;+--+$&$\w_7=w_{10}$\\
$8$&$+;+---$&$w_{8}$&$-;+---$&$\w_8=w_{9}$\\
$9$&$+;-+++$&$w_{9}$&$-;-+++$&$\w_9=w_{8}$\\
$10$&$+;-++-$&$w_{10}$&$-;-++-$&$\w_{10}=w_{7}$\\
$11$&$+;-+-+$&$w_{11}$&$-;-+-+$&$\w_{11}=w_{6}$\\
$12$&$+;-+--$&$w_{12}$&$-;-+--$&$\w_{12}=w_{5}$\\
$13$&$+;--++$&$w_{13}$&$-;--++$&$\w_{13}=w_{4}$\\
$14$&$+;--+-$&$w_{14}$&$-;--+-$&$\w_{14}=w_{3}$\\
$15$&$+;---+$&$w_{15}$&$-;---+$&$\w_{15}=w_{2}$\\
$16$&$+;----$&$w_{16}$&$-;----$&$\w_{16}=w_{1}$\\
\hline
\end{tabular}
\end{center}
\end{table}

The rates constrained by the six conditions~(\ref{six}) now depend on ten independent parameters.
Note that, in~(\ref{six}), the first two equations involve couples of rates related by detailed balance conditions, while the following ones are linear combinations of such couples. 

This space of parameters can be further reduced by fixing the properties of the rate function under spatial symmetry.
For the totally asymmetric dynamics where each spin sees only its East and North neighbours, the rate function is {\it uniquely} determined by the constraint equations~(\ref{six})~\cite{gb2009,cg2013}, up to a scale of time, and is given by~(\ref{rate:ne}). 
Fixing this timescale by the choice $\alpha=2\cosh^2 2 K$, allows to recast~(\ref{rate:ne}) into the compact form:
\beq\label{ku}
w(\s;\{\s_j\})
=\e^{-2 K\s(\s_{E}+\s_{N})}.
\eeq
This form appeared first in~\cite{kun} without the factor 2 in the exponent, which amounts to replacing $\gamma$ by $\tanh K$, and corresponds to a stationary state with the temperature halved.
The uniqueness of this form was first shown in~\cite{gb2009}.

As mentioned above, the rate function~(\ref{rate0}), which corresponds to the case where $V=0$, does not possess the full permutation symmetry of the spins, i.e., is not a function of the local field $\s_E+\s_N+\s_W+\s_S$ only.
It is nevertheless invariant under the spatial left-right and up-down symmetries, which, combined with the global balance condition~(\ref{six}), enforces the condition of detailed balance.

From now on we shall set $\alpha=1$.

\section{Reminder of the dynamics of the directed chain}
\label{sec:remind}

The case of the linear chain will serve as a template for the study of the dynamics of the two-dimensional model.
We therefore begin by a brief reminder of the main results of the study performed in~\cite{cg2011}.

The most general rate function, denoted by $w(\s_n;\{\s_j\})$, where $\s_n$ is the flipping spin and $\{\s_j\}$ is a notation for the configuration of the neighbours $\{\s_{n-1},\s_{n+1}\}$, invariant under spin reversal symmetry, depends on four independent parameters, i.e., takes four values according to the configuration chosen. 
Imposing global balance lowers the number of parameters to three. 
Imposing furthermore the solvability of the dynamical equations (i.e., their linearity) lowers this number to two: a global time-scale parameter $\alpha$, as in~(\ref{rate}), that we set equal to one, and an asymmetry parameter, which, in the particular case of one dimension, encodes entirely the role of irreversibility~\cite{cg2013}.

We thus obtain the following rate function~\cite{cg2011}
\beq\label{direct}
w(\s_n;\{\s_j\})=\frac{1}{2} \Big[1-\gamma\s_n\Big(\frac{1+V}{2}\s_{n-1}+\frac{1-V}{2}\s_{n+1}\Big)\Big ],
\eeq
where $V$, the asymmetry parameter, has the meaning of a velocity,
as can be seen, e.g., by following the motion of domain walls at zero temperature.
This rate function interpolates between the totally symmetric model for $V=0$, which is the usual Ising-Glauber model~\cite{glau}, and the totally asymmetric models for $V=\pm1$.
With this choice of rate function the temporal evolution of the observables defined hereafter obeys linear equations~\cite{cg2011}, as for the case of the Ising-Glauber model~\cite{glau}.

\subsection{Observables}
In order to characterize the relaxation of the system to its stationary state, amongst the various observables which can be monitored, we will focus only on a few of them, either in one or two dimensions.
These are the magnetization,
\beq
M_n(t)=\langle\s_n(t)\rangle,
\eeq
the equal-time correlation function,
\beq
C_n(t)=\langle\s_{0}(t)\s_{n}(t)\rangle,
\eeq
and the two-time correlation function, where $s<t$,
\beq
C_n(s,t)=\langle\s_{0}(s)\s_{n}(t)\rangle.
\eeq
We are also interested in the two-time response, which by translation invariance only depends on the difference
$n-m$,
\beq\label{R}
R_{n-m}(s,t)=\left.\frac{\delta M_n(t)}{\delta H_m(s)}\right\vert_{\{H=0\}},
\eeq
where $H$ is an externally applied field, in reduced temperature units.

\subsection{Symmetric dynamics}

First, for symmetric dynamics ($V=0$), i.e., for the Ising-Glauber model, we have the following results~\cite{glau,gl2000}.

At equilibrium, the characteristic scale of correlations is given by the correlation length $\xi_{\eq}$.
The later is defined by the spatial decay of the spin-spin correlation function at equilibrium
\beq\label{cneq}
C_{n,\eq}=(\tanh K)^{|n|}=\e^{-|n|/\xi_{\eq}}.
\eeq
On the other hand, the temporal decay of the equal-time correlation function to its equilibrium value behaves, according to~\cite{gl2000}, eq.~(3.19), as
\beq\label{cnt}
C_{n,\eq}-C_n(t)\sim\e^{-\alpha t},
\eeq
with
\beq\label{eq:taueq}
\alpha=\frac{2}{\tau_{\eq}},\qquad \tau_{\eq}=\frac{1}{1-\gamma},
\eeq
where $\tau_{\eq}$, the equilibrium relaxation time, is the only characteristic time scale present in the theory.
It is for example the time scale for the decay of the magnetization $M_n(t)=\langle\s_n(t)\rangle$.
Let the initial magnetization be $M_n(0)$.
Then, since all evolution equations are linear in the symmetric case, the magnetization at time $t$ is a discrete spatial convolution
\beq
M_n(t)=M_n(0) * G_n(t)=\sum_{m}M_{n-m}(0)G_m(t),
\eeq
where $G_n(t)$, the magnetization at time $t$ for a random initial condition, with the spin at the origin pointing upwards, that is to say the Green function for the magnetization~\cite{gl2000}, reads\footnote{$I_n$ is the modified Bessel function.}
\beq
G_n(t)=\e^{-t}I_n(\gamma t)\sim \e^{-\alpha_{G}t},
\eeq
with the relaxation rate (inverse relaxation time)
\beq
\alpha_{G}=1-\gamma.
\eeq

For an initially uniformly magnetized system, with $M_n(0)=M$, we have in particular
\beq
M_n(t)=M\e^{-t/\tau_{\eq}}.
\eeq
It is easily shown that $G_n(t)$ is identical to the two-time correlation function,
\beq\label{eq:cn0t}
C_n(0,t)=\langle\s_0(0)\s_n(t)\rangle,
\eeq
which is the overlap between the configuration at time $t$ and the totally disordered
initial condition $\{\s_n(0)\}$.

The two-time equilibrium correlation and response functions decay exponentially with the relaxation rates $\alpha_C$ and $\alpha_R$, both identical to $\alpha_G$,
\beqa
C_{n,\eq}(\tau)=\lim_{s\to\infty}
C_{n}(s,s+\tau)
\sim\e^{-\alpha_{C}\tau},
\label{eq:cneq0}\\
R_{n,\eq}(\tau)=\lim_{s\to\infty}
R_n(s,s+\tau)
\sim\e^{-\alpha_{R}\tau}.
\label{eq:respons}
\eeqa
The relationship between the two quantities is more precisely given by the fluctuation-dissipation theorem~\cite{fdt}
\beq\label{fdstat}
R_{n,\eq}(\tau)=-\frac{\d C_{n,\eq}(\tau)}{\d \tau}.
\eeq
We will refer to $\alpha_C$ and $\alpha_G$ (or $\alpha_R$) as the linear and non linear relaxation rates, respectively.

%%%%%%%%%%%%%%%%%%%%%%%%%%%%%%%%%%%%%%%%%%%%
\begin{figure}[h]
\includegraphics[angle=0,width=1\linewidth]{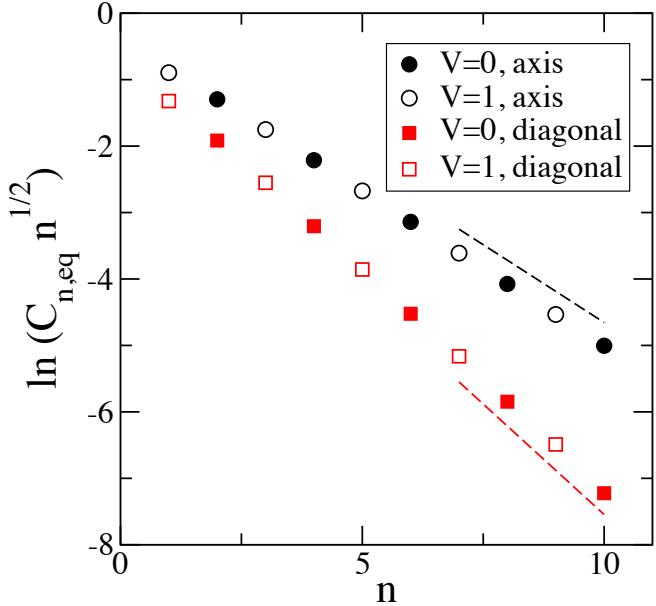}
%\centerline{\epsfxsize=4.25in\ \epsfbox{figure1CG.eps}}
\caption{Stationary spatial correlation function of the two-dimensional model along the axes and diagonals, at $T=3$, for $V=0$ and $V=1$.
The inverse of the slopes yield estimates for 
the equilibrium correlation lengths in axial and diagonal directions. The dashed lines
indicate the inverse of the exactly known values of $\xi_{\eq}$
given in equations (\ref{eq:xi_eq_axis}) and (\ref{eq:xi_eq_diag}). 
The system has been relaxed for $t=1000$ time steps before measuring
the correlation. The data result from averaging over 24\,000 independent runs.
}
\label{fig:C_n_eq}
\end{figure}
%%%%%%%%%%%%%%%%%%%%%%%%%%%%%%%%%%%%%%%%%%%%

%
\subsection{Asymmetric dynamics}

For asymmetric dynamics ($V\ne0$), we have the following results~\cite{cg2011}.

\medskip\noindent (i)
The equal-time correlation function $C_n(t)$ is independent of the asymmetry.
Hence~(\ref{cneq}) and (\ref{cnt}) hold unchanged.
This property, also shared by the spherical model under asymmetric Langevin dynamics~\cite{gl2013}, has its origin in the linearity of the dynamical equations.

\medskip\noindent (ii)
We have
\beq\label{eq:cn0t+}
C_n(0,t)=G_n(t)\sim \e^{-\alpha_{G}t},
\eeq
with a decay rate which varies continuously with $V$,
\beq
\alpha_{G}=1-\gamma\sqrt{1-V^2}.
\eeq
Hence, the larger the asymmetry, the faster the decay: the dynamics is accelerated by the asymmetry.
The complete expression of the Green function is:
\beq\label{greenexact}
G_n(t)=\e^{-t}\left(\frac{1+V}{1-V}\right )^{n/2} I_{n}(\gamma t\,\sqrt{1-V^2}).
\eeq
It is thus invariant under the simultaneous sign changes of $n$ and $V$.
Conversely, for a fixed value of $V$, $G_n(t)$ is spatially asymmetric,
as demonstrated by the expression of the time-independent ratio
\beq\label{GG}
\frac{G_n(t)}{G_{-n}(t)}=\left(\frac{1+V}{1-V}\right )^{n}.
\eeq
We can thus define a length scale associated to this asymmetry by
\beq
\e^{\pm1/\xi_{\mathrm{ asym}}}=\sqrt{\frac{1+V}{1-V}},
\eeq
where the sign $\pm$ corresponds respectively to the cases $V>0$ and $V<0$.
This length, even in $V$, diverges for $V=0$ as $\xi_\mathrm{{asym}}\approx\pm 1/V$ and vanishes for $V=\pm1$.

\medskip\noindent (iii)
At stationarity, the two-time correlation function 
\beq\label{eq:cstat1D}
C_{n,\stat}(\tau)
%=\lim_{s\to\infty}\langle\s_0(s)\s_0(s+\tau)\rangle
=\lim_{s\to\infty}C_{n}(s,s+\tau)
\sim\e^{-\alpha_{C}\tau},
\eeq
is exponentially decreasing with the decay rate
\beq
\alpha_C = \left\{
\begin{array}{lr}
    \alpha_1=1-\gamma\sqrt{1-V^2}=\alpha_G, &(V<V_c)\\
    \alpha_2=V\sqrt{1-\gamma^2},   & (V>V_c).
\end{array}
\right.
\eeq
The two decay rates $\alpha_1$ and $\alpha_2$ are equal at the critical velocity $V_c=\sqrt{1-\gamma^2}$.

On the other hand, the stationary response function reads
\beq
R_{n,\stat}(\tau)=
\sqrt{1-\gamma^2}\,G_n(\tau),
\eeq
implying the equality of relaxation rates $\alpha_R=\alpha_G$.

From now on, unless otherwise stated, we shall focus on the correlation and response functions at coinciding points ($n=0$).

\medskip\noindent (iv)
As a consequence of the above, the fluctuation-dissipation theorem~(\ref{fdstat}) no longer holds.
Its violation can be quantified by the limit stationary fluctuation-dissipation ratio $X_{\stat}$ defined as~\cite{cg2011}
\beq\label{eq:defXstat}
X_{\stat}=\lim_{\tau\to\infty}X_{\stat}(\tau)=\lim_{\tau\to\infty}\frac{R_{0,{\stat}}(\tau)}{-\d C_{0,\stat}(\tau)/\d \tau}.
\eeq

This ratio has the following behaviour:
\begin{enumerate}
\item[(a)]
For $V<V_c$, the relaxation rates of the two quantities are both equal to $\alpha_1$.
It turns out that the sub-leading corrections to the exponential decay are also the same, with the consequence that the stationary fluctuation-dissipation ratio has the finite limit
\beqa\label{X_stat}
 X_{\stat}
&=&\frac{1-\gamma/\sqrt{1-V^2}}{1-\gamma\sqrt{1-V^2}}.
\eeqa
In particular, for $V=0$, we have $X_{\stat}\equiv X_{\eq}=1$, in accordance with~(\ref{fdstat}).

\item[(b)]
For $V>V_c$, this ratio vanishes, because the relaxation rate of the response,
$\alpha_R=\alpha_G=\alpha_1=1-\gamma\sqrt{1-V^2}$, is always larger than the relaxation rate of the correlation,
$\alpha_C=\alpha_2=V\sqrt{1-\gamma^2}$.

\end{enumerate}

In summary, if $V<V_c$, the stationary fluctuation-dissipation ratio has a finite limit at large time because the correlation and response functions have the same asymptotic exponential decay.
For $V>V_c$ the correlation function decays more slowly and $ X_{\stat}=0$.

This dynamical transition has a simple interpretation in terms of the two length scales present in the theory, namely the equilibrium correlation length $\xi_{\eq}$ and the asymmetry length scale $\xi_{\mathrm{ asym}}$.
The condition $V<V_c$ corresponds to the condition $\xi_{\eq}<\xi_\mathrm{ asym}$.
The effect of the asymmetry becomes dominant when $\xi_{\eq}>\xi_\mathrm{ asym}$.
The transition occurs when the two length scales cross~\cite{cg2011}.

\subsection*{Remark}

The limit {\it stationary} fluctuation-dissipation ratio defined in~(\ref{eq:defXstat}) is different from the limit ratio $X_{\infty}$ encountered in the study of the transient regime of nonequilibrium systems (see, e.g., \cite{glX,cala}).
Defining the fluctuation-dissipation ratio as~\cite{X}
\beq\label{Xdef}
X(s,t)=\frac{R_0(s,t)}{\partial C_0(s,t)/\partial s},
\eeq
its asymptotic limit for $\tau=t-s$ large, denoted as $X_{\mathrm{ as}}(s)$~\cite{gl2000}, reads
\beq
X_{\mathrm{ as}}(s)=\lim_{\tau\to\infty}X(s,s+\tau)\equiv
\lim_{t\to\infty}X(s,t),
\eeq
whose limit at large waiting time $s$ yields the limit {\it transient} fluctuation-dissipation ratio
\beq\label{xinfty}
X_{\infty}=\lim_{s\to\infty}X_{\mathrm{ as}}(s)=\lim_{s\to\infty}\lim_{t\to\infty}X(s,t).
\eeq

%%%%%%%%%%%%%%%%%%%%%%%%%%%%%%%%%%%%%%%%%%%%
\begin{figure}[h]
\includegraphics[angle=0,width=1\linewidth]{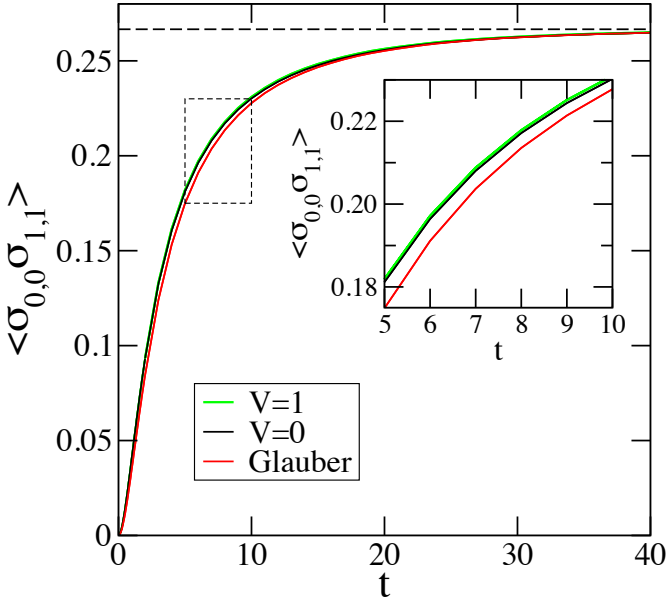}
%\centerline{\epsfxsize=4.25in\ \epsfbox{figure2CG.eps}}
\caption{Growth of the North-East diagonal correlation function $\langle\sigma_{0,0}(t)\sigma_{1,1}(t)\rangle$ for $V=0, V=1$ and for the Glauber case
at temperature $T=3$. 
The horizontal line depicts the exact equilibrium value of the correlation at that temperature. The inset shows a blow-up
of the region within the dashed box of the main figure and illustrates that the correlations grow faster for $V=1$ than for $V=0$, with the Glauber
case yielding the slowest increase.
The data result from averaging over two millions independent runs.
}
\label{fig:corr}
\end{figure}
%%%%%%%%%%%%%%%%%%%%%%%%%%%%%%%%%%%%%%%%%%%%

%%%%%%%%%%%%%%%%%%%%%%%%%%%%%%%%%%%%%%%%%%%%
\begin{figure}[h]
\includegraphics[angle=0,width=1\linewidth]{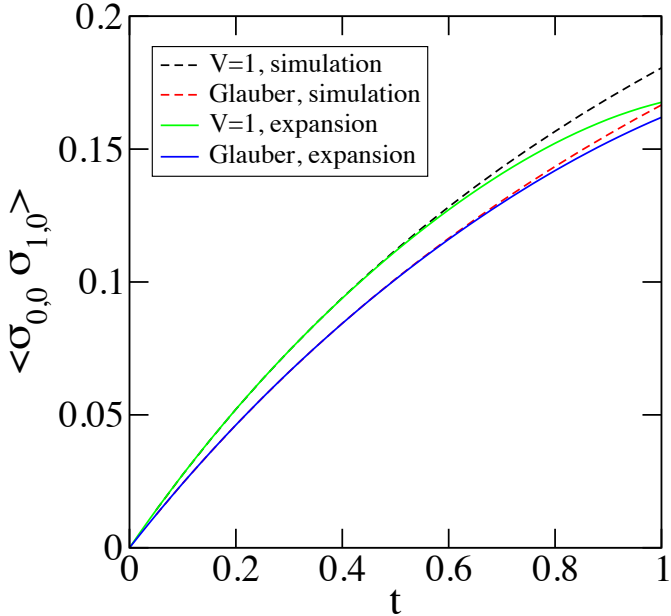}
%\centerline{\epsfxsize=4.25in\ \epsfbox{figure3CG.eps}}
\caption{Comparison of the short-time expansions with simulations for the near neighbour axial correlation function
$\langle \sigma_{0,0}(t)\sigma_{1,0}(t)\rangle$,
both for the Glauber case and for our model with $V=1$. 
\textcolor{black}{(The corresponding curves for $V=0$ are very close to those for $V=1$.)}
Results are shown for the temperature $T=3$.
}
\label{fig:axialshort}
\end{figure}
%%%%%%%%%%%%%%%%%%%%%%%%%%%%%%%%%%%%%%%%%%%%

%%%%%%%%%%%%%%%%%%%%%%%%%%%%%%%%%%%%%%%%%%%%
\begin{figure}[h]
\includegraphics[angle=0,width=1\linewidth]{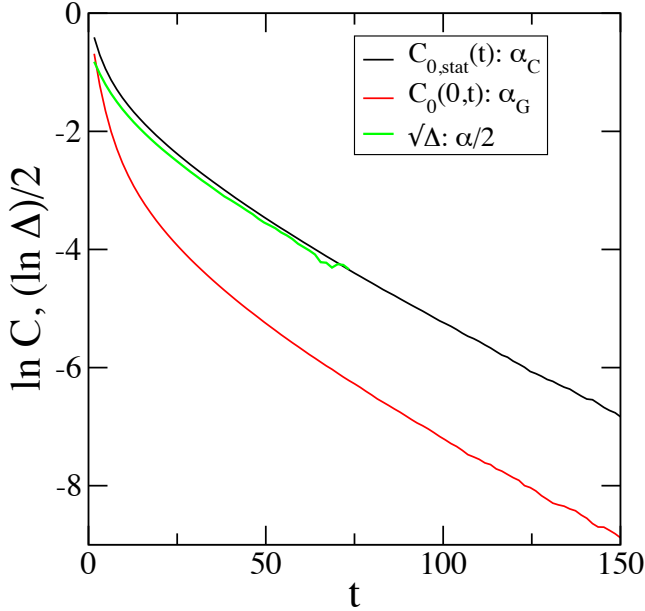}
%\centerline{\epsfxsize=4.25in\ \epsfbox{figure4CG.eps}}
\caption{Comparison of the time dependence of (a): the stationary two-time correlation $C_{0,\stat}(t)$ (\ref{eq:alfaC}), (b): the overlap with the disordered initial condition $C_0(0,t)$ (\ref{eq:alfaG}) and (c):
the approach of the correlation function $\left< \sigma_{0,0}(t) \sigma_{1,1}(t) \right>$ to its stationary value. 
The green line shows the square-root of $\Delta(t)$ (see~(\ref{eq:diff}))
in order to demonstrate that the exponential decays indicate the relationships $\alpha/2\simeq \alpha_C\simeq\alpha_G$, see main text.
All three cases correspond to $V=0$ and temperature $T=3$.
For $C_{0,\stat}(t)$ we averaged over 16\,000 independent runs, whereas an average of much more runs has been done
for the other two quantities: 660\,000 runs for $C_0(0,t)$ and two millions run for $\Delta(t)$.
}
\label{fig:alpha}
\end{figure}
%%%%%%%%%%%%%%%%%%%%%%%%%%%%%%%%%%%%%%%%%%%%

%%%%%%%%%%%%%%%%%%%%%%%%%%%%%%%%%%%%%%%%%%%%
\begin{figure}[h]
\includegraphics[angle=0,width=1\linewidth]{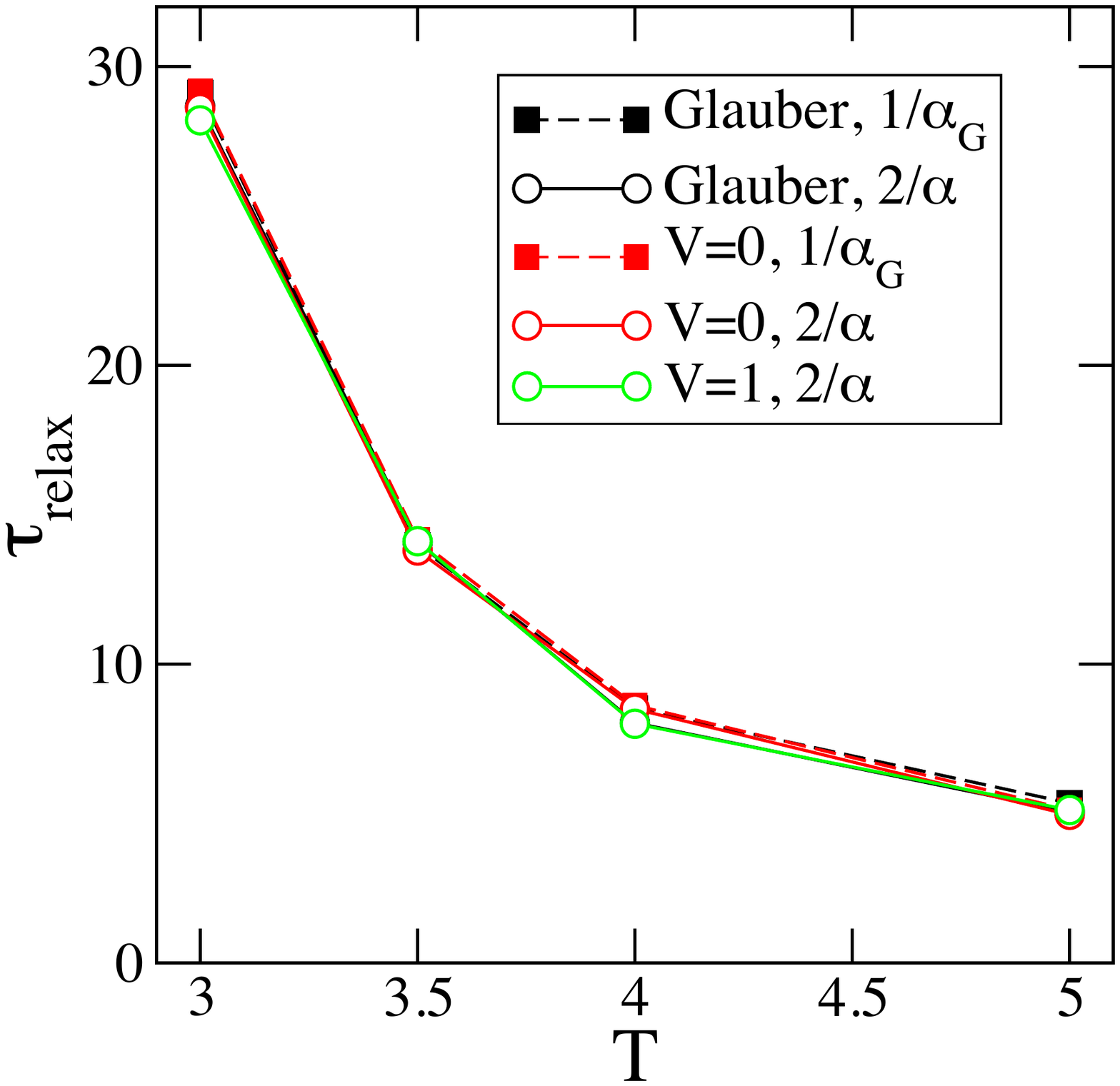}
%\centerline{\epsfxsize=4.25in\ \epsfbox{figure5CG.eps}}
\caption{Comparison of the relaxation times obtained from the exponential decays of~(\ref{eq:diff}) and~(\ref{eq:alfaG}), for $V=0$ and Glauber, and~(\ref{eq:diff}) for $V=1$.
We thereby use the relationship $\tau_{\rel} = 1/\alpha_G$ for (\ref{eq:alfaG})  and $\tau_{\rel} =2/\alpha$ for (\ref{eq:diff}). Error bars are comparable to the sizes of the symbols.}
\label{fig:tau_eq}
\end{figure}
%%%%%%%%%%%%%%%%%%%%%%%%%%%%%%%%%%%%%%%%%%%%

%
%
%
\section{Dynamics in the paramagnetic phase: equal-time correlations}
\label{sec:bulk1}
We start our study of the two-dimensional directed Ising model on the square lattice by the investigation of the equal-time correlation function in the paramagnetic phase $T>T_c$\footnote{The critical temperature $T_c=2/\ln(1+\sqrt{2})\approx 2.269185$ is such that $\sinh 2/T_c=1$.}.

We already know that its stationary value is independent of the asymmetry introduced
by a finite velocity $V$, since the rate~(\ref{rate}) leads to a Gibbsian stationary measure.
The question is therefore whether, during its temporal evolution, this correlation function has, or has not, a dependence on the velocity.
In one dimension, as recalled in section~\ref{sec:remind}, the equal-time correlation is independent of the velocity and this is to be related to the fact that the dynamical equations are linear.
In two dimensions the dynamics is non linear, and therefore, as can be expected, and as will be demonstrated below, the equal-time correlation bears a dependence on the velocity during its temporal evolution.

In practice, in order to elucidate the dynamics of our model, we relied on two approaches:
\begin{enumerate}
\item We performed extensive Monte Carlo simulations where we used the rate (\ref{rate}) in order to decide whether a proposed
flip of the spin $\sigma_{m_i,n_i}$, located at site $i$ with integer coordinates $(m_i,n_i)$,
should be accepted. Typically, we simulated systems composed of $300 \times 300$ spins.
In all cases, a large number of independent runs were done and ensemble averages were determined (see
the figure captions for details on the number of samples used in our study).
\item We also carried out short-time expansions of the observables.
The method used is straightforward, see, e.g., ref.~\cite{wang}.
\end{enumerate}

\subsection{Equilibrium equal-time correlation function}

As mentioned above, at stationarity, the equal-time correlation function
does not depend on $V$, and its
asymptotic behaviour is known~\cite{coy}. 
This therefore can be used as a test to validate our algorithm.
In figure \ref{fig:C_n_eq} we compare results obtained for $V=0$ and $V=1$ in two different directions, 
respectively along the $x-$axis, and along the North-East diagonal, with
\beqa
C_{n,\eq}({\rm axis}) = \left< \sigma_{0,0} \sigma_{0,n} \right>,
\nonumber\\
C_{n,\eq}({\rm diag}) = \left< \sigma_{0,0} \sigma_{n,n} \right>,
\eeqa
both at the same temperature $T=3$. 
As expected, the results do not depend on the value
of $V$. 
The asymptotic expression for the static correlation function for $T>T_c$ (see eqs.~(2.43) and (2.46)
in chapter XI of \cite{coy}) is of the form:
\begin{equation}
C_{n,\eq} \sim \e^{- \left| n \right|/\xi_{\eq}}/n^{1/2}~,
\end{equation}
where the equilibrium correlation length $\xi_{\eq}$ depends on the angle. 
This behaviour is the two-dimensional counterpart of~(\ref{cneq}).
Along the axes one has that
\begin{equation}
\xi_{\eq}({\rm axis})^{-1}=\ln \frac{1-v}{v(1+v)}~,
\label{eq:xi_eq_axis}
\end{equation}
where $v=\tanh 1/T$, whereas along the diagonals
\begin{equation}
\xi_{\eq}({\rm diag})^{-1} = 2 \ln \frac{1-v^2}{2 v}~.
\label{eq:xi_eq_diag}
\end{equation}
Therefore, the inverse of the equilibrium correlation length is identical to the slope when
plotting $\ln \left( C_{n,\eq} n^{1/2} \right)$ as a function of $n$, see figure \ref{fig:C_n_eq}.
Our results obtained from simulations using the rate (\ref{rate}) agree perfectly with the exact results,
indicated by the dashed lines in figure \ref{fig:C_n_eq}.

\subsection{Relaxation of the equal-time correlation function}

For $T>T_c$, when the system is initially out of equilibrium, the equal-time correlation function converges exponentially fast to its stationary (equilibrium) value, and this holds for any value of $V$.

Let us first focus on the near-neighbour North-East diagonal correlation function 
$\langle\sigma_{0,0}(t)\sigma_{1,1}(t)\rangle$,
whose equilibrium expression is known exactly \cite{coy,mon62} (see Appendix). 
Figure~\ref{fig:corr} depicts the behaviour of this function for $V=0$ and $V=1$.
For comparison we also plot the same function obtained with the Glauber rate.
The difference between the curves corresponding to $V=0$ and $V=1$ is of the order of $10^{-3}$ and therefore hardly visible on this figure.
The existence of a difference can be demonstrated by looking at the short-time expansion of the correlation function.
We could obtain such an expansion at order 4 in time $t$ with little computing effort:
\beq
\langle\sigma_{0,0}(t)\sigma_{1,1}(t)\rangle\approx 
a_0+a_1 t+\frac{a_2}{2!}t^2+\frac{a_3}{3!}t^3+\frac{a_4}{4!}t^4,
\eeq
with $a_0=0$ since the initial condition is disordered, and
\beqa
a_1=0,\quad
a_2=2\gamma^2,\quad
a_3=-2(4\gamma^2+3\gamma^4),
\nonumber\\
a_4=24\gamma^2+60\gamma^4+\frac{29}{2}\gamma^6+\frac{7}{2}V^2\gamma^6.
\label{eq:coefV}
\eeqa
The role of the velocity is thus made apparent.
Its importance increases as temperature is lowered, and vanishes at infinite temperature.

For comparison we give the corresponding short-time expansion for the Glauber case:
\beqa
a_1=0,\quad
a_2=8a^2,\quad
a_3=-32a^2,\quad
\nonumber\\
a_4=4\left(24a^2+64a^4-160a^3b+58a^2b^2-30ab^3+3b^4\right),
\label{eq:coefGlau}
\eeqa
where
\beq
a=\frac{\gamma (2+\gamma ^2)}{4(1+\gamma ^2)},
\qquad
b=\frac{\gamma ^3}{4(1+\gamma ^2)}
\eeq
are the parameters entering the Glauber rate~(\ref{glau2D}), when it is written as an expansion on a basis of spin operators, as 
\beqa
w(\s;\{\s_{j}\})=\frac{1}{2}\left[1-a\,\s(\s_{E}+\s_{N}+\s_{W}+\s_{S})\right.
\nonumber\\
\left.
+b\,\s(\s_{E}\s_{N}\s_{W}+\s_{N}\s_{W}\s_{S}+\s_{W}\s_{S}\s_{E}+\s_{S}\s_{E}\s_{N})\right].
\label{eq:glauber}
\eeqa
At high temperature, the expansions in $\gamma$ of the coefficients~(\ref{eq:coefV}), for the $V=0$ case, and~(\ref{eq:glauber}), for the Glauber case, match at leading order.

We also performed the short-time expansion of the North-West diagonal correlation function $\langle\sigma_{0,0}(t)\sigma_{-1,1}(t)\rangle$ in order to test the possible presence of an anisotropy in the temporal evolution of the correlation function, even when $V=0$.
We obtained:
\beqa
a_1=0,\quad
a_2=2\gamma^2,\quad
a_3=-2\left(4\gamma^2+3\gamma^4\right),
\nonumber\\
a_4=4\left(6\gamma^2+15\gamma^4+4\gamma^6\right).
\eeqa
At this order this correlation function does not depend on $V$ and the difference between the North-East and North-West correlations only appears in $a_4$.
It is equal to $(7V^2-3)\gamma^6 t^4/2$, which changes sign when $V$ increases.
The dependence on $V$ of the North-West correlation function only appears at order 5:
\beq
a_5=-64\gamma^2-352(\gamma^4+\gamma^6)+2V^2\gamma^6-\frac{1}{2}(99+V^2)\gamma^8.
\eeq

For completeness we also performed the short-time expansion of the axial near neighbour correlation function
$\langle\sigma_{0,0}(t)\sigma_{1,0}(t)\rangle$ for our model:
\beqa
a_1=\gamma,\quad
a_2=-\left(2\gamma+\gamma^3\right),
\nonumber\\
a_3=\frac{1}{2}\left(8\gamma+18\gamma^3+3\gamma^5+V^2\gamma^5\right),
\nonumber\\
a_4=-\frac{1}{4}\left(32\gamma+168\gamma^3+156\gamma^5+24V^2\gamma^5+9\gamma^7+3V^2\gamma^7\right),
\eeqa
and for the Glauber case:
\beqa
a_1=2a\quad
a_2=-4a
\nonumber\\
a_3=8\left(a+5a^3-5a^2b+ab^2\right),
\nonumber\\
a_4=-16\left(a+15a^3-15a^2b+5ab^2\right).
\eeqa
Again, the two expansions match at high temperature.
Figure~\ref{fig:axialshort} depicts a comparison between the two models at very short times.

As a conclusion, both the axial and diagonal near neighbour correlation functions depend on the velocity.
The correlations grow faster for $V$ larger. 
Let us emphasize that these differences, hardly visible for $T>T_c$, are on the contrary easily seen at low temperature~\cite{ustocome}.
The largest observed differences are nevertheless between the correlations for our model and for the Glauber case.
Finally the presence of an anisotropy, even for $V=0$, is revealed by the difference between the diagonal North-East and North-West correlations.

\subsection{Approach of the correlation function to its stationary value}

The approach of the near neighbour diagonal correlation function to its stationary value can be characterized by the decay rate $\alpha$ of the difference
\beq\label{eq:diff}
\Delta(t)=\langle \sigma_{0,0}\sigma_{1,1}\rangle-\langle \sigma_{0,0}(t)\sigma_{1,1}(t)\rangle
\sim\e^{-\alpha t}
\eeq
which is the counterpart of~(\ref{cnt}).
As seen in the previous subsection, $\Delta(t)$ depends only weakly on $V$.

Anticipating on the sequel,
in figure~\ref{fig:alpha},  we compared the rate $\alpha$ with the other characteristic relaxation rates appearing in the dynamics of the system, for the case $V=0$.
Inspired by the known behaviour of the linear chain, we compared the decay~(\ref{eq:diff}) to that of $C_{0,\stat}(\tau)$ and $C_0(0,t)$, respectively given by~(\ref{eq:alfaC}) and~(\ref{eq:alfaG}) below.
This comparison indicates that, for the case $V=0$, $\alpha/2\simeq \alpha_C= \alpha_G$, paralleling the 
property~(\ref{eq:taueq}).

As a conclusion, in the paramagnetic phase, the equal-time correlation function of the two-dimensional directed Ising model behaves almost as its one-dimensional counterpart, since its dependence on the velocity is only very weak.
While the decay rate $\alpha$ is weakly dependent on the velocity, this is not the case of the decay rates $\alpha_C$ and $\alpha_G$, as analyzed in the next section.

\section{Two-time observables}
\label{sec:bulk2}

As discussed in section 3, the study of the one-dimensional directed Ising model revealed for a given temperature
the existence of a critical velocity $V_c$ that separates a regime of weak violation of the fluctuation-dissipation
theorem for $V < V_c$ from one of strong violation for $V > V_c$. 
In the subsections below, following what was done in one dimension, we study different
quantities that all should display signatures of this transition, provided it also takes place in the two-dimensional
directed Ising model. 

\subsection{Determination of the different relaxation rates}

%%%%%%%%%%%%%%%%%%%%%%%%%%%%%%%%%%%%%%%%%%%%
\begin{figure}[h]
\includegraphics[angle=0,width=1\linewidth]{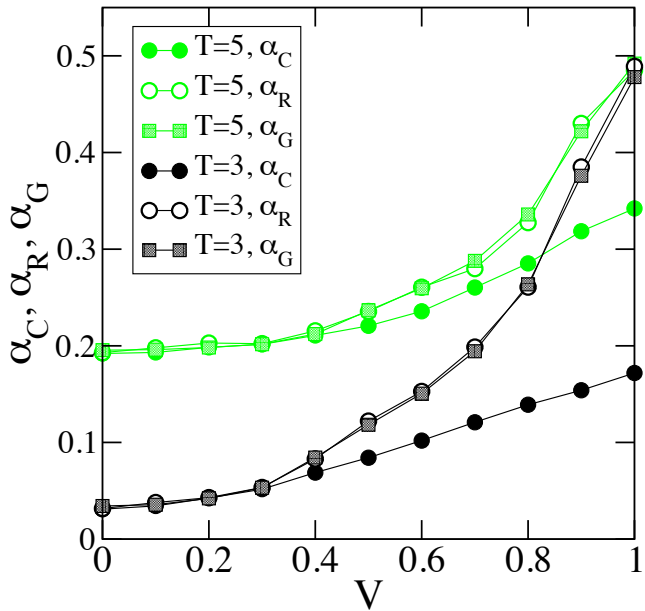}
%\centerline{\epsfxsize=4.25in\ \epsfbox{figure6CG.eps}}
\caption{Decay exponents $\alpha_C$, $\alpha_R$, and $\alpha_G$ obtained from the
stationary correlation $C_{0,\stat}(\tau)$, the stationary integrated response function $\chi_{TRM}(\tau)$,
and the correlation $C_0(0,t)$ far from stationarity. 
The condition $\alpha_R > \alpha_C$ for $V > V_c$
gives an estimate for the critical velocity shown in figure \ref{fig:Vc}. 
}
\label{fig:alphaTous}
\end{figure}
%%%%%%%%%%%%%%%%%%%%%%%%%%%%%%%%%%%%%%%%%%%%

We start by investigating different relaxation rates which give a measure of the characteristic time scales.
These relaxation rates are obtained from the exponential decay of various, stationary and non-stationary, correlation
and response functions, whose definitions follow from the one-dimensional case and that we repeat here for convenience.

\begin{enumerate}

\item 
From the stationary two-time correlation,
\beq\label{eq:alfaC}
C_{0,\stat}(\tau)
=\lim_{s\to\infty}\langle\s_{0,0}(s)\s_{0,0}(s+\tau)\rangle
\sim\e^{-\alpha_C \tau},
\eeq
we obtain the linear relaxation time $1/\alpha_C$, i.e., the decay time of stationary fluctuations.
This quantity is the two-dimensional counterpart of~(\ref{eq:cstat1D}).

\item
The overlap with the disordered initial condition yields the correlation
\beq\label{eq:alfaG}
C_0(0,t)=\langle\s_{0,0}(0)\s_{0,0}(t)\rangle\sim\e^{-\alpha_G t}
\eeq
from which the nonlinear relaxation time $1/\alpha_G$ can be determined.
This quantity is the two-dimensional counterpart of~(\ref{eq:cn0t}) and~(\ref{eq:cn0t+}).

\item
Finally, the stationary response
\beq\label{respons}
R_{0,\stat}(\tau) = \lim\limits_{s \to \infty} R_0(s,s+\tau)\sim\e^{-\alpha_R \tau}
\eeq
is the counterpart of~(\ref{eq:respons}) and yields the additional relaxation time $1/\alpha_R$.
\end{enumerate}

Figure \ref{fig:tau_eq} shows that the relaxation times obtained from the exponential decays of~(\ref{eq:diff}) and~(\ref{eq:alfaG})
are identical for Glauber dynamics and for the $V=0$ case. 
Using the relationship $\tau_{\rel} = 1/\alpha_G$ for (\ref{eq:alfaG})  and $\tau_{\rel} =2/\alpha$ for (\ref{eq:diff}),
we find that the different quantities yield the same value for the relaxation time. The same $\tau_{\rel}$ prevails
for $V=1$ when looking at the approach of the near neighbour diagonal correlation function to its
stationary value.
%%%%%%%%%%%%%%%%%%%%%%%%%%%%%%%%%%%%%%%%%%%%
\begin{figure}[h]
\includegraphics[angle=0,width=1\linewidth]{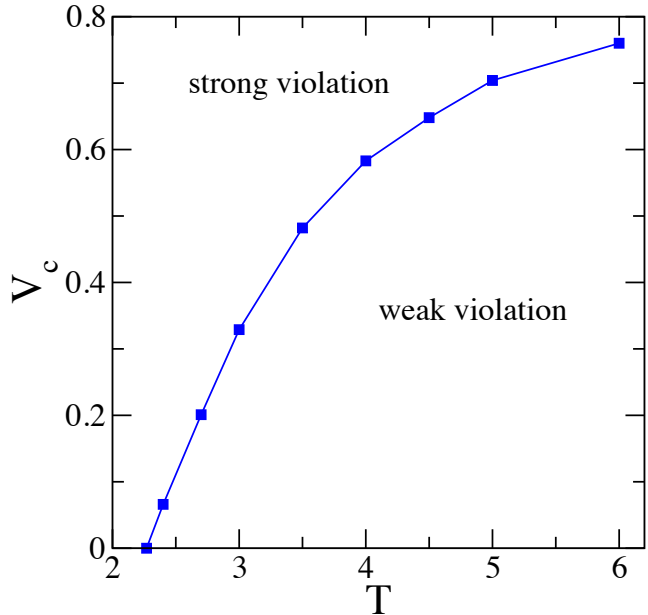}
%\centerline{\epsfxsize=4.25in\ \epsfbox{figure7CG+.eps}}
\caption{Critical value of the velocity $V_c$ as a function of temperature. 
The blue squares are estimates
for $V_c$ obtained from the condition $\xi_{\asym} = \xi_{\eq}$, where $\xi_{\asym}$ is the length scale
describing the asymmetry of the diagonal correlator $C_n(0,t)$, whereas $\xi_{\eq}$ is the equilibrium
correlation length of the two-dimensional Ising model, see figure \ref{fig:xi_asym}.
}
\label{fig:Vc}
\end{figure}
%%%%%%%%%%%%%%%%%%%%%%%%%%%%%%%%%%%%%%%%%%%%

In figure \ref{fig:alphaTous} we show the dependence on $V$ of the different rates $\alpha_C$, $\alpha_G$, and $\alpha_R$ for two different temperatures.
We note that a scenario similar to that of the one-dimensional case prevails.
On the one hand, the rate $\alpha_C$ is found to be equal to $\alpha_R$ for $V<V_c$, after which it varies approximately linearly with $V$. 
The rate $\alpha_G$, on the other hand, is found to be equal to $\alpha_R$.
These later rates correspond both to quantities which measure the memory of a perturbation applied on the system far away in the past, and therefore we infer their strict equality.
The velocity $V_c$, where the values of $\alpha_C$ and $\alpha_R$ start to differ,
indicates the transition between the two dynamical regimes of violation of the
fluctuation-dissipation theorem. 
This can be used to construct the $T-V_c$ phase diagram.
In practice, however, it is rather difficult to reliably determine $V_c$ in that way. 
We will see in the
following that using the asymmetry correlation length provides a much more reliable way for the determination of the critical velocity, as shown in figure~\ref{fig:Vc}. 

\subsection*{Remark}

For convenience as well as in order to achieve better statistics,
we do not determine directly the response $R_{0,\stat}(\tau)$~(\ref{respons}), but instead we calculate time integrated quantities.
We consider two different protocols, similar to those used routinely in studies of systems relaxing towards
a steady state \cite{HenPlebook}. 
In the first protocol,
a field with amplitude $H_0$ (in reduced temperature units) is applied between times $s'$ and $s$ after having reached the steady state.
This yields the dimensionless time integrated susceptibility
\beq
\chi_{TRM}(s,s+\tau) = \int_{s'}^{s}\d u \, R_0(u,s+\tau),
\eeq
which is proportional to the thermoremanent magnetization.
At stationarity it decays with the same relaxation time as the response itself.
In the second protocol,
the field is applied 
continuously after time $s$, and the response is measured for times larger than $s$, yielding the dimensionless susceptibility
\beq
\chi_{ZFC}(s,s+\tau) =  \int_{s}^{s+\tau} \d u\, R_0(u,s+\tau),
\eeq
proportional to the zero field cooled magnetization.
Practically, following~\cite{Bar98}, we apply at each site $n$ a spatially random field $H_n = H_0 \epsilon_n$,
where $\epsilon_n = \pm 1$, following one of the two protocols just discussed, and compute the staggered magnetization
\beq
M =\frac{1}{ N} \overline{\Big\langle \sum_n \epsilon_n\sigma_n  \Big\rangle},
\eeq
from which the susceptibility $\chi=M/H_0$ is obtained.
For the data obtained with the TRM protocol and discussed in the following, we applied the field for 100
time steps after having reached the steady state and then measured the susceptibility. 
We carefully checked that 
our results do not depend on the number of time steps during which the field has been switched on.

\subsection{Determination of the asymmetry correlation length}

In order to determine the asymmetry correlation length we proceed as for the one-dimensional case (see~(\ref{GG}))
and make the ansatz
\beq
\frac{C_{n}(0,t)}{C_{-n}(0,t)}=\e^{-2 n/\xi_{\asym}},
\label{eq:C_ratio}
\eeq
where $C_n(0,t)=\langle\s_{0,0}(0)\s_{n,n}(t)\rangle$ is the overlap with the totally random initial state.
Note that this assumes
that the length $\xi_{\asym}$ is independent of time, which is confirmed by our simulations.

%%%%%%%%%%%%%%%%%%%%%%%%%%%%%%%%%%%%%%%%%%%%
\begin{figure}[h]
\includegraphics[angle=0,width=1\linewidth]{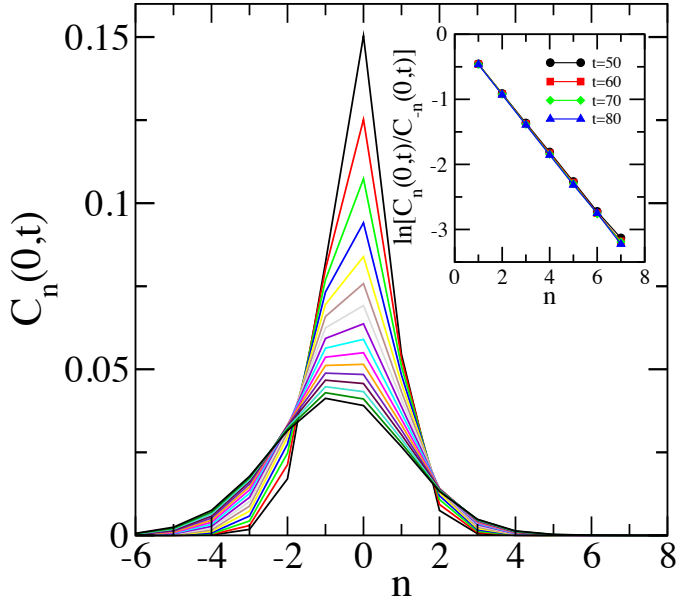}
%\centerline{\epsfxsize=4.25in\ \epsfbox{figure8CG.eps}}
\caption{
Correlation function $C_n(0,t)$
for $T=2.4$ and $V=0.1$. 
The value of $t$ increases from 5 to 20 from top to bottom. 
The data result from averaging over 160\,000 independent runs. Inset:
The asymmetry length scale follows from the exponential change with $n$ of the ratio $C_n(0,t)/C_{-n}(0,t)$.
}
\label{fig:C_n}
\end{figure}
%%%%%%%%%%%%%%%%%%%%%%%%%%%%%%%%%%%%%%%%%%%%

\begin{figure}[h]
\includegraphics[angle=0,width=1\linewidth]{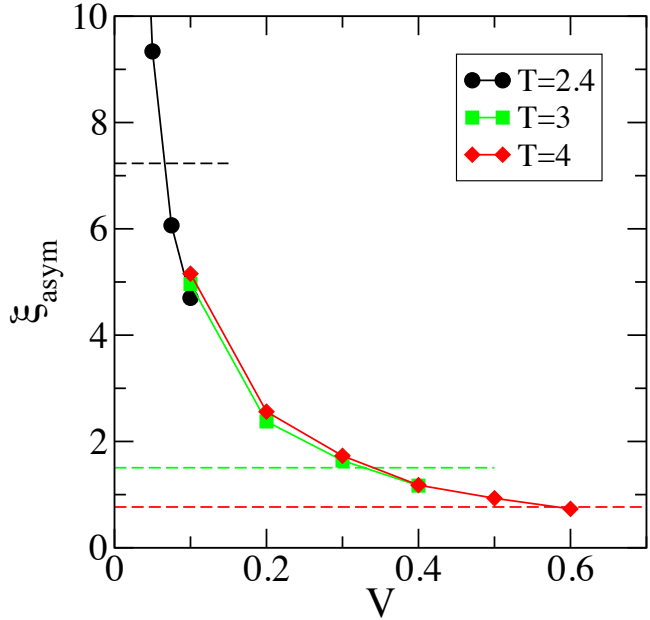}
%\centerline{\epsfxsize=4.25in\ \epsfbox{figure9CG.eps}}
\caption{Asymmetry length scale $\xi_{\asym}$ as a function of $V$ for three different temperatures.
The dashed lines indicate the equilibrium correlation lengths of the two-dimensional Ising model at
the different temperatures. 
The boundary between the weak and strong violation regimes is obtained 
when these two length scales coincide: $\xi_{\asym} = \xi_{\eq}$.
}
\label{fig:xi_asym}
\end{figure}

The typical behavior of $C_{n}(0,t)$ is shown in figure \ref{fig:C_n} for $T=2.4$
and $V = 0.1$, where we consider the correlation in the diagonal direction, i.e.,
$C_{n}(0,t) = \langle \sigma_{0,0}(0) \sigma_{n,n}(t) \rangle$. 
The expected asymmetry is very notable. Plotting the ratio (\ref{eq:C_ratio}) as a function of
$n$ in a log-log plot, see the inset of figure \ref{fig:C_n}, then
allows to determine a value for $\xi_{\asym}$ from the slope. 
Results from this procedure are shown in 
figure \ref{fig:xi_asym}. 
This asymmetry correlation length can then be compared to the
equilibrium correlation length $\xi_{\eq}$ (see~(\ref{eq:xi_eq_diag}) and the dashed
lines in the figure). 
This allows to determine the critical velocity $V_c$,
which is obtained by the condition $\xi_{\asym} = \xi_{\eq}$. 
The critical line $V_c(T)$ obtained in this way, as depicted in figure~\ref{fig:Vc}, 
is in fair agreement, at least for smaller values of $T$, 
with the line obtained from the relaxation times. 
In fact, using $\xi_{\asym}$ yields a more precise determination of $V_c$, via the
intersection of the two lines $\xi_{\asym}$ and $\xi_{\eq}$.

%%%%%%%%%%%%%%%%%%%%%%%%%%%%%%%%%%%%%%%%%%%%
%%%%%%%%%%%%%%%%%%%%%%%%%%%%%%%%%%%%%%%%%%%%

%
\subsection{Fluctuation-dissipation ratio at stationarity}

We close with a discussion of the fluctuation-dissipation ratio at stationarity. 
As discussed previously for the one-dimensional case, the violation of the 
fluctuation-dissipation theorem in the directed system can show up in two different ways. 
Indeed,
for $V < V_c$ the limit fluctuation-dissipation ratio $X_{\stat}$ is expected to be larger than zero
but smaller than the equilibrium value $X_{\eq} = 1$. 
This regime corresponds to a weak violation of the fluctuation-dissipation theorem~\cite{cg2011}. 
In contrast, for $V > V_c$, one should be in the regime of strong violation of the fluctuation-dissipation theorem,
characterized by $X_{\stat}=0$.

%%%%%%%%%%%%%%%%%%%%%%%%%%%%%%%%%%%%%%%%%%%%
\begin{figure}[h]
\includegraphics[angle=0,width=0.6\linewidth]{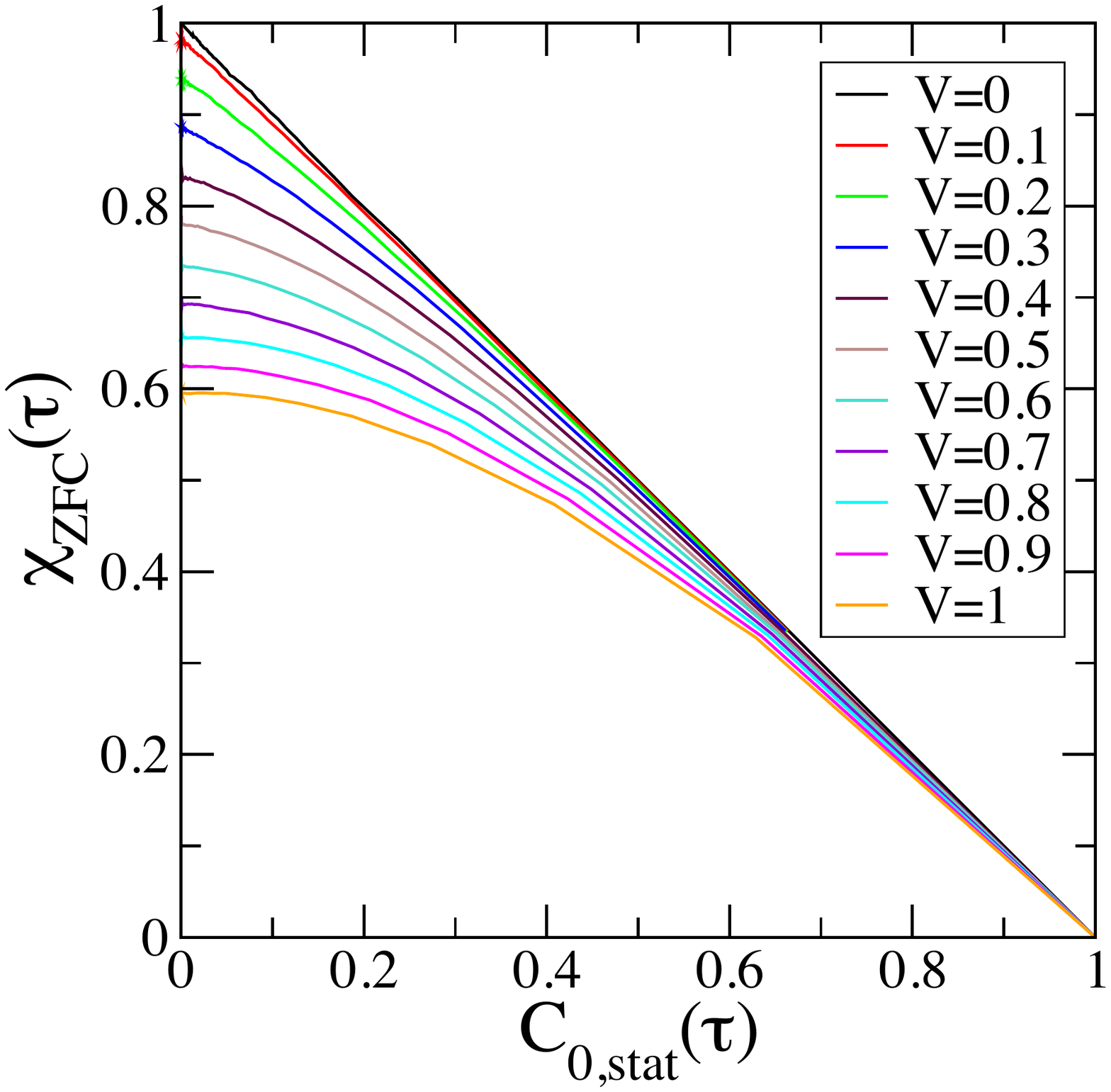}
\includegraphics[angle=0,width=0.6\linewidth]{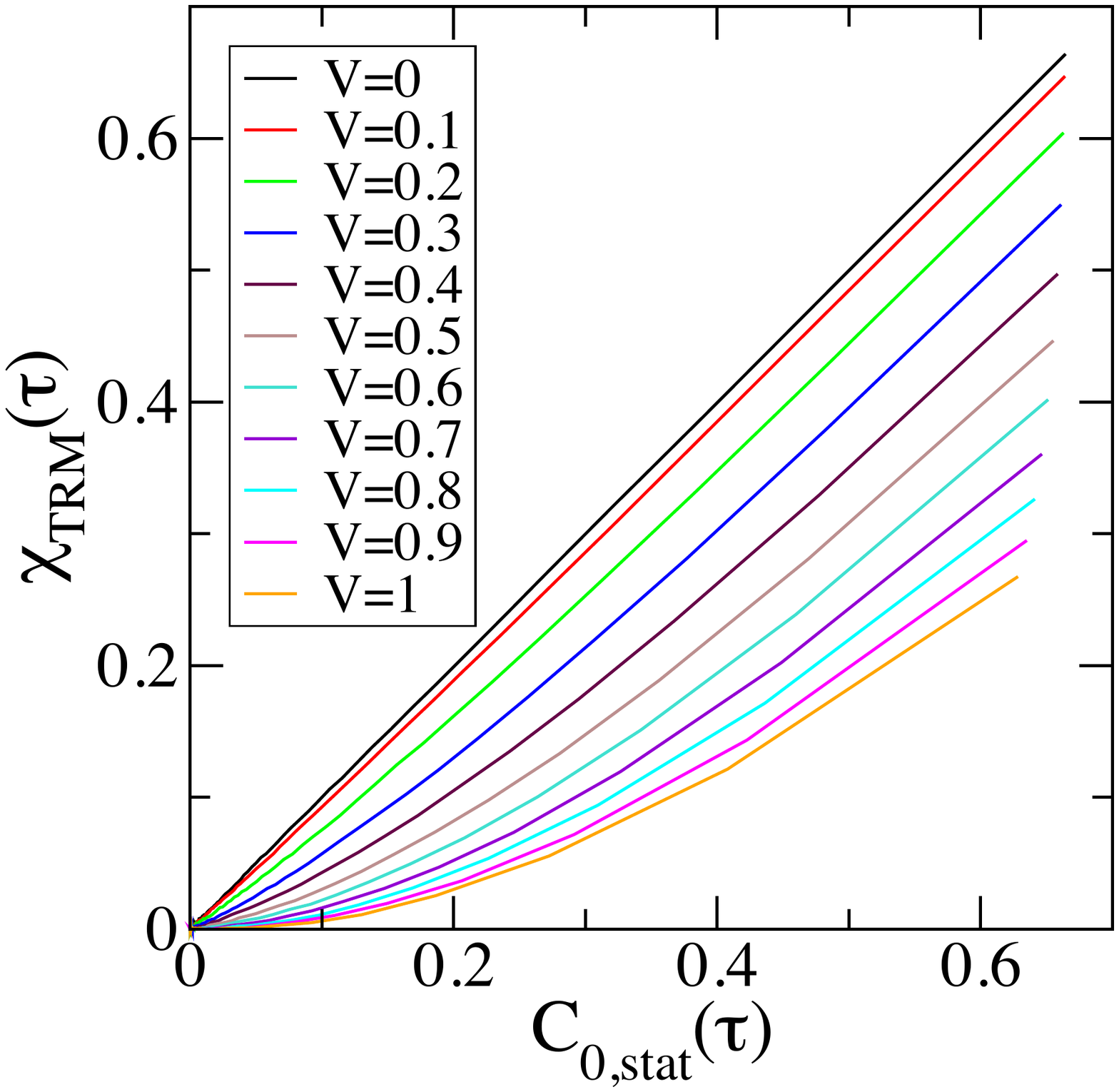}
%\centerline{\epsfxsize=2.25in\ \epsfbox{figure10CG.eps}
%\epsfxsize=2.25in\ \epsfbox{figure10MCG.eps}}
\caption{Stationary integrated response functions plotted against the stationary correlation at
the temperature $T=3$ for different velocities $V$. 
The system is relaxed for $s=1\,000$ MCS before computing
the quantities as a function of $\tau = t-s$. We carefully checked that we are indeed at stationarity,
i.e., that the data only depend on the time difference $\tau$. 
For the correlation function the data result
from averaging over $1\,000$ independent runs. For the response function we average over typically
$100\,000$ runs.
Left panel: zero-field cooled susceptibility, right panel: thermoremanent susceptibility.
}
\label{fig:chivsC}
\end{figure}
%%%%%%%%%%%%%%%%%%%%%%%%%%%%%%%%%%%%%%%%%%%%

In figure \ref{fig:chivsC} the stationary integrated susceptibilities
are plotted against the stationary autocorrelation $C_{0,\stat}$ for $T=3$. 
Figure \ref{fig:chivsC}a shows the zero-field cooled susceptibility $\chi_{ZFC}$
as a function of $C_{0,\stat}$.
This time integrated response increases with $\tau = t-s$.
At the same time,
the autocorrelation decreases with $\tau$. 
Inspection of the curves $\chi_{ZFC}(C_{0,\stat})$ indeed reveals that different behaviors are encountered
for small and high velocities. 
For small values of $V$, $\chi_{ZFC}$ exhibits
a finite slope when $C_{0,\stat}$ is small, indicating the weak violation of the fluctuation-dissipation theorem. For large values of $V$,
however, $\chi_{ZFC}$ has zero slope when $C_{0,\stat}$ is small, which is the signature of the
strong violation of the fluctuation-dissipation theorem. 

%%%%%%%%%%%%%%%%%%%%%%%%%%%%%%%%%%%%%%%%%%%%
\begin{figure}[h]
\includegraphics[angle=0,width=1\linewidth]{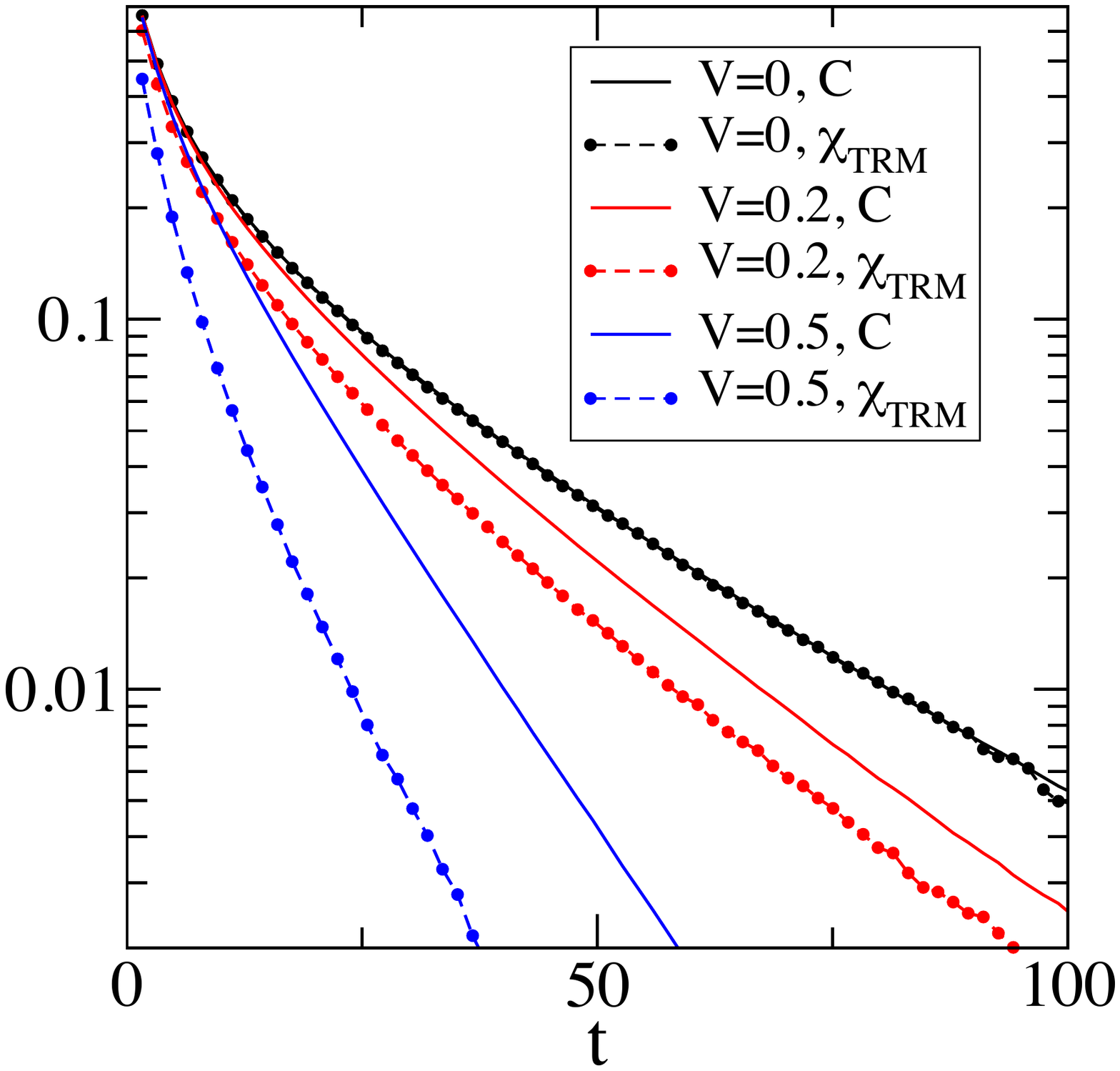}
%\centerline{\epsfxsize=4.25in\ \epsfbox{figure11CG.eps}}
\caption{The correlation and the stationary thermoremanent magnetization for $T=3$ and three different velocities: $V=0$, $V=0.2<V_c$ and $V=0.5>V_c$, where $V_c\approx 0.34$ (see figure~\ref{fig:xi_asym}).
For the computation of the stationary thermoremanent magnetization, a magnetic field of strength $h_0 = 0.05$ has been switched
on for 100 time steps after having reached stationarity. 
The number of independent realizations are the same as in figure \ref{fig:chivsC}.
}
\label{fig:C_M}
\end{figure}
%%%%%%%%%%%%%%%%%%%%%%%%%%%%%%%%%%%%%%%%%%%%

%%%%%%%%%%%%%%%%%%%%%%%%%%%%%%%%%%%%%%%%%%%%
\begin{figure}[h]
\includegraphics[angle=0,width=1\linewidth]{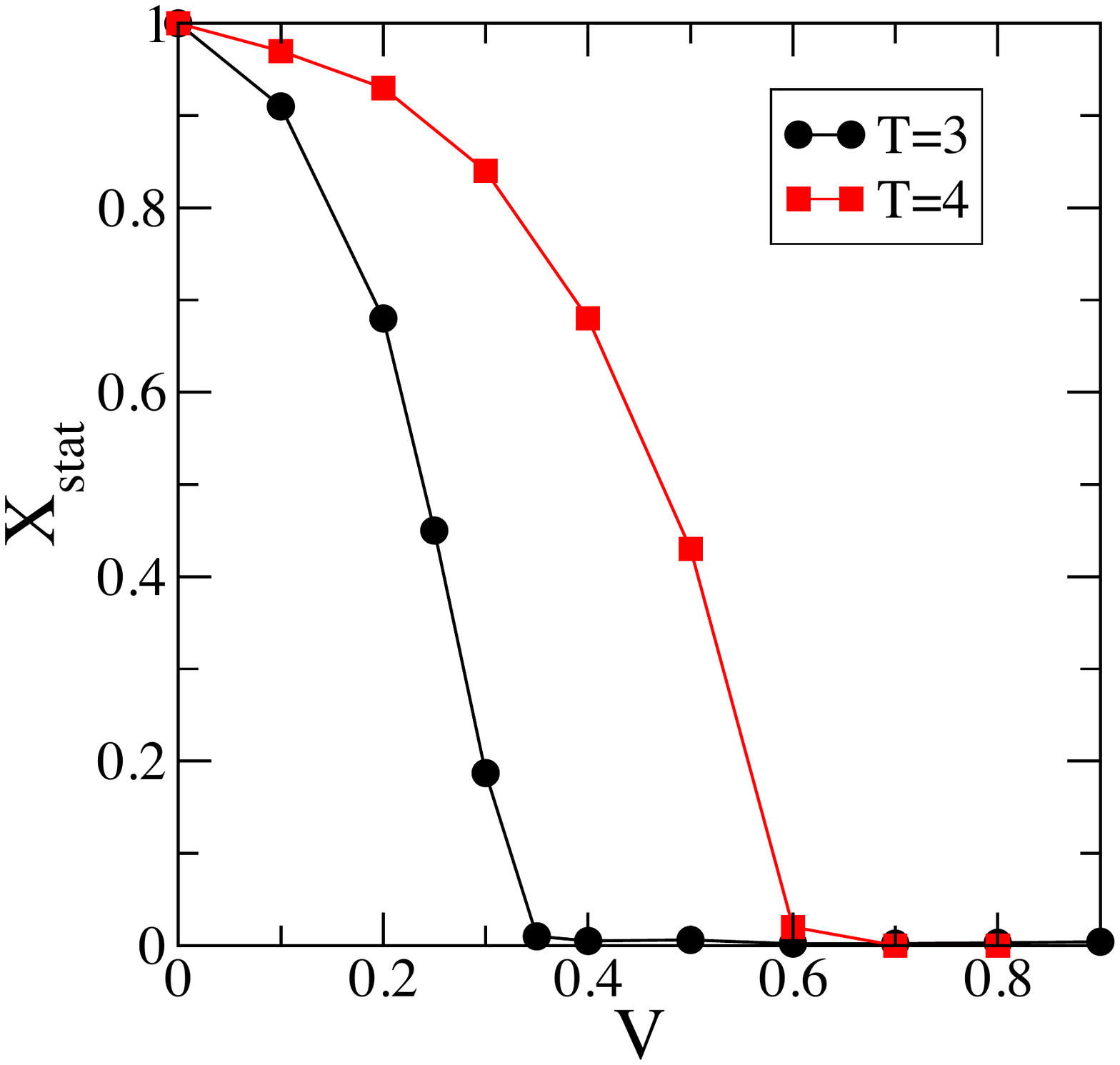}
%\centerline{\epsfxsize=4.25in\ \epsfbox{figure12CG.eps}}
\caption{The stationary limit fluctuation-dissipation ratio as a function of velocity $V$ for two different 
temperatures. These data are obtained from comparing the stationary integrated response function 
with the stationary correlation, see figure \ref{fig:C_M} for an example.
}
\label{fig:Xstat}
\end{figure}
%%%%%%%%%%%%%%%%%%%%%%%%%%%%%%%%%%%%%%%%%%%%

The limit stationary fluctuation-dissipation ratio $X_{\stat}$ is best extracted from the thermoremanent susceptibility.
The ratio of the decaying susceptibility
$\chi_{TRM}(\tau)$ to the correlation $C_{0,\stat}(\tau)$ then yields $X_{\stat}$ for $\tau$ large, see figure \ref{fig:chivsC}b.
In figure \ref{fig:C_M} we show the result of this analysis for three different velocities. 
For $V =0$, $C_{0,\stat}$ and $\chi_{TRM}$ are identical, in agreement with the fluctuation-dissipation theorem. 
For $V < V_c$, see the curves for $V=0.2$, $\chi_{TRM}$ and $C_{0,\stat}$ are no longer identical. 
Still, they decay with the same decay constant $\alpha_R =\alpha_C$, yielding, after an early-time regime, parallel lines in a log-log plot and a value for $X_{\stat}$ that is between 0 and 1. 
Finally, for $V > V_c$, see the curves
for $V=0.5$, $\alpha_R \neq \alpha_C$, and the ratio of $\chi_{TRM}$ and $C_{0,\stat}$ goes to zero for large $\tau$. 
Using the thermoremanent magnetization therefore provides us with a straightforward way to determine the limit fluctuation-dissipation ratio $X_{\stat}$. 
The result of this
procedure, shown in figure \ref{fig:Xstat}, indeed reveals the two different dynamical regimes.

\section{Conclusion}

In the present work we have pursued the investigation of the role of spatial asymmetry
and irreversibility on the dynamics of spin systems, by considering the two-dimensional ferromagnetic
Ising model on the square lattice with asymmetric dynamics. 
This line of research, initiated in~\cite{gb2009}, was followed by several works~\cite{cg2011,cg2013,gl2013}. 
In particular ref.~\cite{cg2011} is devoted to the case of the linear Ising chain with asymmetrical dynamics, while ref.~\cite{gl2013} is devoted to the spherical model with asymmetric linear Langevin dynamics.

The outcomes of the present study turn out to share
many common features with those of~\cite{cg2011,gl2013}.
Indeed, we show the existence of two regimes of violation of the fluctuation-dissipation theorem in the nonequilibrium stationary state: when the asymmetry parameter is less than a threshold value, there is weak violation of the fluctuation-dissipation theorem, with a finite asymptotic 
fluctuation-dissipation ratio, whereas this ratio vanishes above the threshold.
The present results suggest that this novel kind of dynamical transition in nonequilibrium stationary states might be quite general.

While for the two models mentioned above, the directed Ising chain~\cite{cg2011} and the ferromagnetic spherical model with asymmetric linear Langevin dynamics~\cite{gl2013},
the equal-time spin-spin correlation function does not feel the presence of the asymmetry in the dynamics,
for the two-dimensional Ising case studied in the present work, this correlation function bears a dependence on the asymmetry.
For the two former models the dynamics obeys linear equations, while for the model studied in the present work this no longer holds.

As was already the case for the linear chain, one of the consequences of the introduction of an asymmetry in the dynamics is the acceleration of the relaxation process towards stationarity.
This can be seen for example on figure~\ref{fig:alphaTous}, where the relaxation rates, either linear or non linear, increase drastically with the asymmetry parameter $V$.

Some recent papers discuss how, for Markovian systems, the convergence towards the equilibrium Boltzmann-Gibbs distribution is accelerated by choosing dynamical rates violating detailed balance but fulfilling the condition of global balance~\cite{manou,ren,suwa,turi,fern,saka,ichi}.
There is therefore a conceptual overlap between those references and our work as well as the previous references~\cite{gb2009,cg2011,cg2013,gl2013}, but the former are more interested in algorithmic issues, while in our work as well as in~\cite{gb2009,cg2011,gl2013,cg2013} the concern is oriented towards the role of irreversibility on the physical properties of the models.
It would nevertheless be interesting to deepen the connections between the two topics.

In a companion paper, we shall address the situation encountered at criticality and in the ferromagnetic phase of the two-dimensional Ising model with asymmetric dynamics.

\ack We wish to thank J-M Luck for fruitful discussions and M Henkel for his interest at an early stage of the present study.
MP acknowledges financial support by the US National
Science Foundation through grant DMR-1205309.

\appendix

\section{Equilibrium properties of the two-dimensional Ising model}
In this appendix we recall some classical results on the near-neighbour correlation functions of the equilibrium Ising model on the square lattice used in the bulk of the paper (see e.g.~\cite{coy,mon62}).
Let
\beq
k_{>}=\sinh^2 2/T=k_{<}^{-1}.
\eeq
Then, along the $x-$axis:
\begin{itemize}
\item 
for $T<T_c$, 
\beq
\langle\s_{0,0}\s_{0,1}\rangle=\sqrt{1+k_{<}}\left(\frac{1}{2}+\frac{1-k_{<}}{\pi}K(k_{<}^2)\right),
\eeq
\item
for $T>T_c$:
\beq
\langle\s_{0,0}\s_{0,1}\rangle=\sqrt{\frac{1+k_{>}}{k_{>}}}\left(\frac{1}{2}+\frac{k_{>}-1}{\pi}K(k_{>}^2)\right).
\eeq

\end{itemize}

\noindent
Along the diagonal:
\begin{itemize}
\item 
for $T<T_c$, 
\beq
\langle\s_{0,0}\s_{1,1}\rangle=\frac{2}{\pi}E(k_{<}^2),
\eeq
\item
for $T>T_c$:
\beq
\langle\s_{0,0}\s_{1,1}\rangle=\frac{2}{\pi k_{>}}\left(E(k_{>}^2)+(k_{>}^2-1)K(k_{>}^2)\right).
\eeq
\end{itemize}

In these expressions
\beq
K(z)=\int_{0}^{\pi/2}\d x\,\frac{1}{(1-z\sin^2 x)^{1/2}}
\eeq
is the elliptic function of the first kind, while
\beq
E(z)=\int_{0}^{\pi/2}\d x\,(1-z\sin^2 x)^{1/2}
\eeq
is the elliptic function of the second kind.

\end{document}